\documentclass[showpacs,preprintnumbers,english,twocolumn,amsmath,amssymb]{revtex4-1}
\usepackage[T1]{fontenc}
\usepackage[latin1]{inputenc}
\usepackage{graphicx}
\usepackage{amssymb}
\usepackage{verbatim}
\usepackage{epic,eepic}
\usepackage{babel}

\begin{document}

\title{Density and Temperature of  Fermions from Quantum Fluctuations}

\author{ Hua Zheng$^{a,b)}$ and Aldo Bonasera$^{a,c)}$}
\affiliation{
a)Cyclotron Institute, Texas A\&M University, College Station, TX 77843, USA;\\
b)Physics Department, Texas A\&M University, College Station, TX 77843, USA;\\
c)Laboratori Nazionali del Sud, INFN, via Santa Sofia, 62, 95123 Catania, Italy.}




\begin{abstract}
A novel  method to determine the density and temperature of a system is proposed based on quantum fluctuations typical of Fermions in the limit where the reached temperature T is small
compared to the Fermi energy $\epsilon_f$ at a given density $\rho$.  Quadrupole and particle multiplicity fluctuations relations are derived in terms of $\frac{T}{\epsilon_f}$. This method is valid for infinite and finite fermionic systems, 
 in particular we apply it to heavy ion collisions using the Constrained Molecular Dynamics (CoMD) approach which includes the Fermi statistics.  A preliminary comparison to available experimental data is discussed as well.
We stress the differences with  methods based on classical approximations. The derived 'quantum' temperatures are systematically lower than the corresponding  'classical' ones.  With the proposed method we may get important informations 
on the Equation of State (EOS) of quantum Fermi systems  to order O($\frac{T}{\epsilon_f})^3$, in particular near the Liquid-Gas (LG) phase transition and at very low densities where quantum effects are dominant.
\end{abstract}


\maketitle

In recent years, the availability of heavy-ion accelerators which provide colliding nuclei from a few MeV/nucleon to GeV/nucleon and  new and performing $4\pi$ detectors, has fueled a field of research loosely referred to as Nuclear Fragmentation. The characteristics of the fragments produced 
depend on the beam energy and the target-projectile combinations which can be externally controlled \cite{1,2,10}.
Fragmentation  experiments could provide informations about the nuclear matter properties and constrain the EOS of nuclear matter\cite{Csernai}.
 From conventional nuclear physics we know that there is a stable equilibrium state at the normal nuclear density $\rho_0=0.145-0.17 fm^{-3}$
  with a compressibility  in the range of $K=180-240$ MeV  and a binding energy of  15-16 MeV/nucleon  \cite{1,2,10,Csernai,shlomo}. 
   Even though a large variety of experimental data and refined microscopic models exist, to date it does not exist a method to determine densities and temperatures reached during the collisions, which takes into account the genuine quantum nature
  of the system.
In this work we discuss some properties at finite temperatures assuming either a classical gas or a quantum Fermi system.  We show
that at the densities and temperatures of interest the classical approximation is not valid.  This is at variance with many experimental and theoretical results in heavy ion collisions near the Fermi energy  \cite{10,albergo,pocho,15,15b,michela,15a} which assume the classical approximation to be valid. 
We base our method on fluctuations estimated from an event by event determination of fragments arising  after the energetic collision.  A similar method has recently been applied to observe suppression of 
fluctuations in a trapped Fermi gas\cite{prl}. We go beyond the method of \cite{prl} by including quadrupole fluctuations as well to have a direct measurement of densities and temperatures for subatomic systems for which it is difficult to obtain such informations in a direct way.
 We also suggest a method for calculating an excitation energy which should minimize collective effects and could be applied when a limited information is available,  for example if only  light cluster  are measured. We apply the proposed method to microscopic CoMD approach \cite{17}
  which includes Fermionic statistics. The resulting densities and temperatures calculated using protons and neutrons, even though surprising at first,  give a perfectly reasonable EOS and a clear first order phase transition.  
\begin{figure}
\centering
\includegraphics[width=0.9\columnwidth]{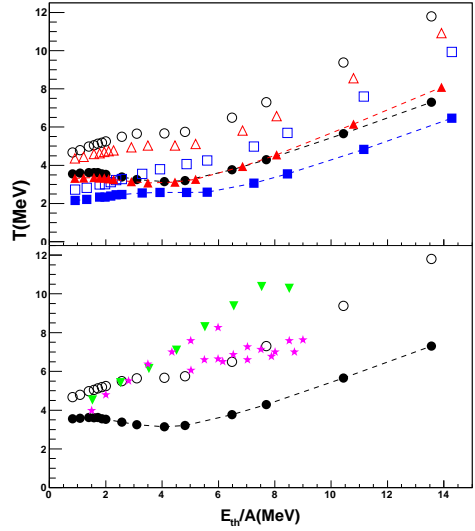} 
\caption[]{Temperature versus thermal energy per particle derived from quantum fluctuations (full symbols joined by  dashed lines) compared to the classical case (open symbols). (Top) Circles refer to protons, squares to neutrons and triangles to protons and neutrons.
(Bottom) Same as above for protons. Data: down triangles from classical quadrupole fluctuations \cite{15b}, star symbols from particle ratios \cite{15}. }
\label{Fig1}
\end{figure}

A method for measuring the temperature was proposed in  \cite{15b} based on momentum fluctuations of detected particles.  A quadrupole $Q_{xy}=<p^2_x-p^2_y>$ is defined in a direction transverse to the beam axis (z-axis) and the average is performed, for a given particle type, over events. Such a quantity is zero
in the center of mass of the equilibrated emitting source.
Its  variance  is given by the simple formula:
\begin{equation}
 \sigma^2_{xy}=\int d^3p(p^2_x-p^2_y)^2f(p)
\label{1}
\end{equation}
where f(p) is the momentum distribution of particles.  In \cite{15b} a classical Maxwell-Boltzmann distribution of particles at temperature $T_{cl}$ was assumed which gives: $ \sigma^2_{xy}=\bar N 4m^2T_{cl}^2$,  m is the mass of the fragment. $\bar N$ is the average number of particles which could be conveniently normalized to one. 
 In heavy ion collisions,  the produced particles  do  $\it not$  follow
classical statistics thus the correct distribution function must be used in eq.(1).  Protons(p), neutrons(n), tritium etc. follow the Fermi statistics while, deuterium, alpha etc., even though they are constituted of nucleons, should follow the Bose statistics.  In this work we will concentrate on fermions only and in particular
p and n which are abundantly  produced in the collisions  thus carrying important informations on the densities and temperatures reached.  

Using a Fermi-Dirac distribution f(p) and expanding to  $O(T/\epsilon_f)^4$, where  $\epsilon_f=\epsilon_{f0}(\frac{\rho}{\rho_0})^{2/3}=
36(\frac{\rho}{\rho_0})^{2/3}$ MeV is the Fermi energy of nuclear matter,  we get \cite{8}:
\begin{equation}
 \sigma^2_{xy}=\bar N[\frac{16m^2\epsilon^2_f}{35}(1+\frac{7}{6}\pi^2(\frac{T}{\epsilon_f})^2)+O(\frac{T}{\epsilon_f})^4]
\label{2}
\end{equation}
  This result is in evident contrast with the classical one:
even at zero T and ground state density $\rho_0$, quadrupole fluctuations arise from the Fermi motion, but those fluctuations cannot be observed since in this case the number of emitted particles  $\bar N=0$.    The quadrupole fluctuations depend
on temperature and density through $\epsilon_f$, thus we need more informations in order to be able to determine both quantities. 

 Within the same framework we can calculate the fluctuations of the p,n multiplicity distributions.  These are given by \cite{8}:
\begin{equation}
 \frac{<(\Delta N)^2>}{\bar N}=\frac{3}{2}\frac{T}{\epsilon_f}+O(\frac{T}{\epsilon_f})^3
\label{3}
\end{equation}
This is a very simple result valid in the indicated approximation.  The difference with the classical case is again striking (the ratio in eq.(3) equal to one for a classical perfect gas).  This relation has also been derived and applied to trapped Fermi gases in \cite{prl}.
These quantities, eqs.(2-3),  can be easily verified experimentally and the corresponding densities and temperatures can be evaluated for each physical situation.
  \begin{figure}
\centering
\includegraphics[width=0.8\columnwidth]{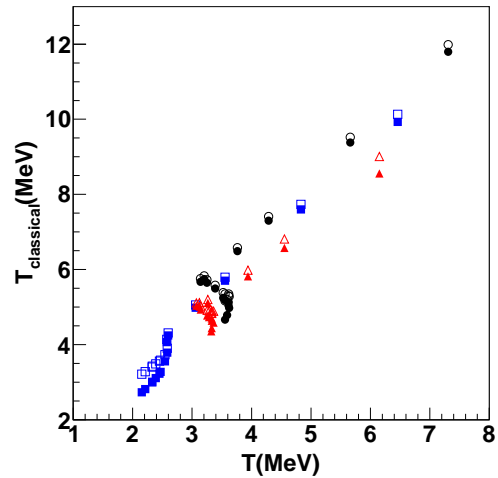}
\caption[]{Classical temperatures vs quantum temperatures. Symbols as in figure 1, open symbols refer to Bauer's approximation, eq.(6).}
\label{Fig2}
\end{figure} 

  To illustrate the strength of our approach we simulated
$^{40}Ca+^{40}Ca$ heavy ion collisions at fixed impact parameter $b=1fm$ and beam energies  $E_{lab}/A$  ranging from 4 MeV/A up to 100 MeV/A.  Collisions were followed up to a maximum time 
$t=1000 fm/c$ in order to accumulate enough statistics.  Particles emitted at later times (evaporation) could affect somehow the results and this might be important especially at the lowest beam energies.
The choice of central collisions was dictated by the desire to obtain full equilibration.  This however, did not occur especially
at the highest beam energies due  to a partial transparency for some events.  For this reason the quadrupole in the transverse direction, eq.(1), was chosen.  Furthermore, in order to correct for collective effects as much as possible, we defined a 'thermal' energy as:
%
\begin{equation}
\langle \frac{E_{th}}{A}\rangle=\frac{E_{cm}}{A}-[\langle\frac{E_{p(n)}}{\bar N_{p(n)}}\rangle-\frac{3}{2}\langle \frac{E_{p(n)xy}}{\bar N_{p(n)}}\rangle]-Q_{value}
\label{4}
\end{equation}
where $\langle\frac{E_{p(n)}}{\bar N_{p(n)}}\rangle$ and $\langle \frac{E_{p(n)xy}}{\bar N_{p(n)}}\rangle$ are the average total and transverse kinetic energies (per particle) 
of protons (and/or neutrons). $Q_{value}=\frac{\bar N_{p(n)}}{Z(N)}8 MeV$, similarly for  protons plus neutrons.  8 MeV is the average binding energy of a nucleon, Z (N) the total charge (neutron number) 
of the system and $\bar N_{p(n)}$
the average number of protons (neutrons) emitted at each beam energy. 
 Clearly for a completely equilibrated system all the center of mass energy, $\frac{E_{cm}}{A}$,  is converted into thermal energy (plus the $Q_{value}$).  If not, our approximation
will account for some corrections, and this will become more and more exact when many fragment types are included in eq.(4).  However, this approximation might be important in experiments where only some fragment types are detected or  if, because of the time evolution of the system, different particles are sensitive
to different excitation energies, for instance if some particles are produced early or late in the collision.

In figure 1 (top) we plot the estimated temperatures at various 'thermal' energies both for the quantum (full symbols) and classical approximations (open symbols).  As we see the quantum case is systematically lower than the classical one.
We also notice a difference if the T are estimated from the proton distributions (circles), or neutrons (squares) or the sum of the two (triangles).  This is clearly a Coulomb effect which gets smaller as expected at higher energies as we will demonstrate more in detail below.
   The back-bending observed at $T\approx 3MeV$ for all cases indicates a liquid-gas phase transition, in particular
we observe that such a back-bending is more  marked for the protons case as first discussed by Gross  \cite{gross}.  In the bottom part of figure 1, we compare the protons results to experimental data. The down triangles are derived using the 'classical' quadrupole fluctuations \cite{15b} thus should be
very similar to our classical results and the agreement is reasonable at the lowest excitation energies.  However, we stress that the experimental data were obtained for  different systems at a fixed 35MeV/A beam energy.  In particular projectile like fragments (PLF) were isolated and analyzed and the excitation energy
was obtained from all fragments differently from eq.(4).  Thus there might be a mismatch in the abscissa and this could be especially important for large excitations.  Also the detector acceptance might be important.  Similar  considerations apply to the data \cite{15} obtained using double particle ratios (star symbols) \cite{albergo}. 
Even though in the latter case it  is perfectly correct to use classical approximations since different particle types are used (usually a mixture of bosons and fermions \cite{albergo}), the underlying assumption is that all those particles are sensitive to the same density and temperature.  If T and $\rho$ 'seen' by different particles are different, then the results give  some kind of 'averaging'
which might hide interesting quantum effects.  In the top part of figure 1 we see that temperatures are different for protons and neutrons at a given excitation energy,  thus we expect that other particles might give different T.  This implies that different  particle ratios might produce different results \cite{15}. 

Using eqs.(1-2), we can easily show that, in the region of validity,  the 'classical' $T_{cl}$ is always larger than the 'quantum' temperature T:
\begin{equation}
T_{cl}=\sqrt{\frac{4\epsilon_f^2}{35}+\frac{2\pi^2}{15}T^2}
\label{5}
\end{equation}
A similar result has been found by Bauer\cite{bauer} in order to explain the large 'apparent' temperature observed in particles spectra. In \cite{bauer} a relation between the final (classical) temperature $T'_{cl}$ and the input Fermi-Dirac T was found:
\begin{equation}
T'_{cl}\approx \frac{2\epsilon_f}{5}[1+\frac{5\pi^2}{12}(\frac{T}{\epsilon_f})^2]
\label{6}
\end{equation}
The ratio $\frac{T}{\epsilon_f}$ entering the equations above can be directly obtained from eq.(3). Even though eqs.(5-6) might look different at first sight, they give very similar results as can be seen in figure 2  where the classical $T_{cl}$ is plotted vs. the quantum one. 
  Bauer's approximation, eq.(6),  is given by the open symbols.   Thus we argue that quantum temperatures are smaller than derived when fitting experimental results with a classical approximation.  The reason of such small quantum temperatures is
 the Fermi energy entering eq.(5) or (6). In passing, we notice that our approximation, eq.(5),
reproduces better the numerical results plotted in fig.2 of ref.\cite{bauer}.

    \begin{figure}
\centering
\includegraphics[width=0.8\columnwidth]{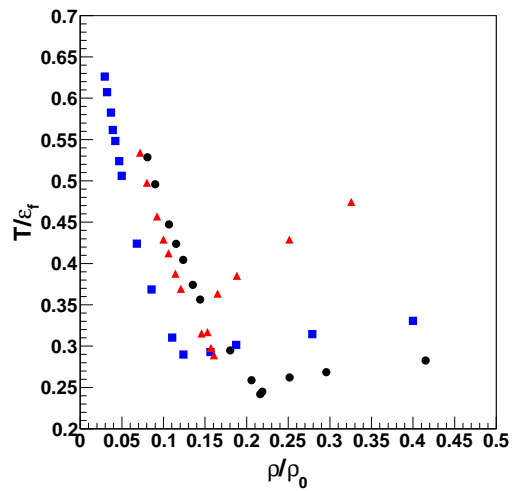} 
\caption[]{Temperature divided the Fermi energy versus density normalized to the ground state one derived from quantum fluctuations, eqs.(1-3). Symbols as in figure 1.}
\label{Fig3}
\end{figure}  
In figure 3 we plot the ratio $\frac{T}{\epsilon_f}$  directly obtained from eq.(3), versus reduced density which is obtained from eqs.(2) and (3). The lowest T (highest $\frac{T}{\epsilon_f}$) corresponds to the lowest beam energy as well and gives the lowest density, especially for the neutrons case.
  This result might be surprising at first, but it simply tells us that at the lowest energies nucleons from the surface of the colliding nuclei come into contact.  Those nucleons are located in a low density region, especially neutrons which do not feel  the Coulomb field.
   With increasing beam energy, the overlapping region increases and more and more fermions are emitted. At about 20 MeV/A a large number of nucleons are excited and the emission from surface becomes a volume emission.
   This explain the minimum in the plot, which is due to the increase of T and $\epsilon_f$ when deeper regions of the nuclei are affected.  Fragmentation starts around the beam energy  which gives the minimum in the plot, where we observe a power law in the mass distribution as well.
The lowest density (as well as T) is explored by the neutrons only.
    It is important to stress that the ratio plotted in figure 3  is always smaller than one which confirms the approximations used in eqs.(1-6). 
    
   The best way to visualize the results is by plotting the energy density $\epsilon=\langle \frac{E_{th}}{A}\rangle \rho$ versus temperature as in figure 4.  Now different particle types scale especially at high T where Coulomb effects are expected to be small. A rapid variation of the energy density is 
    observed around $T\approx 2MeV$ for neutrons and  $T\approx 3MeV$ for the other cases 
    which indicates a first order phase transition\cite{michela}. Notice that  a 'plateau' in the caloric curve i.e. $\langle \frac{E_{th}}{A}\rangle$ vs T  \cite{15,pocho} has been experimentally observed around $T_{cl}\approx 6MeV$. Such a value agrees with our classical approximation plotted in figure 1, but differs greatly 
    with the quantum results, figure 4.  It would be very interesting to reanalyze the experimental data in the framework of this paper. We also notice that Coulomb effects become negligible at around $T=3MeV$ where the phase transition occurs.  The smaller role of the Coulomb field in the
    phase transition has recently been discussed experimentally in the framework of the Landau's description of phase transitions \cite{me}.  
    
 \begin{figure}
\centering
\includegraphics[width=0.8\columnwidth]{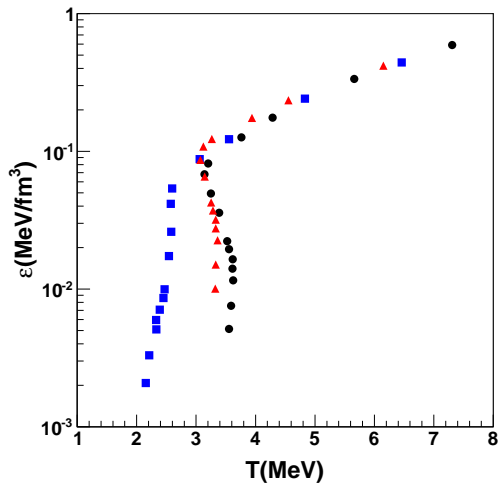} 
\caption[]{Energy density versus temperature. Symbols as in figure 1.}
\label{Fig4}
\end{figure} 
In conclusion, in this work we have addressed a general method for deriving densities and temperatures of fermions.  The method has a validity $O(\frac{T}{\epsilon_f})^3$ and higher order terms might be included if needed. 
In the framework of the  Constrained Molecular Dynamics model, which includes Fermi statistics,  
we have discussed collisions of heavy ions
below 100MeV/A and obtained densities and temperatures at each bombarding energy. Knowing the thermal energy of the system, we can derive the energy density and temperature reached during the collision.  We have been able to bridge low energy phenomenology, i.e.
particles evaporation from the surface, with the fragmentation of the system.  Because of its general validity the approach could be applied to any fermionic system but for temperatures below the corresponding Fermi energies. We are also thinking about trapped Fermi gases \cite{prl} where the complete EOS could
be derived following our method. Our approach is completely at variance with previous ones based on classical mechanics.
The results we have obtained here in a model case confirms that the classical approximation is unjustified.  The tools we have   proposed can be easily generalized to other fermion types, tritons, helions etc., and a comparative study of the EOS for different particles  will be very interesting.  We have seen 
in this work that different particles like neutrons and protons explore different density and temperature regions. Open problems such as Mott transitions, pairing etc. in low density matter 
might be addressed through a detailed study of the EOS using different fermions.  A more conclusive study could be achieved if Boson-like particles could be included
in the approach.  This aspect will be the goal of our future work. 
\begin{acknowledgments} We thank prof. J.Natowitz for 'animated' discussions.
\end{acknowledgments}

\end{document}